\documentclass{SciPost}

\hypersetup{
    colorlinks,
    linkcolor={red!50!black},
    citecolor={blue!50!black},
    urlcolor={blue!80!black}
}

\usepackage[bitstream-charter]{mathdesign}
\usepackage{amsmath}

\usepackage{amssymb}

\usepackage{amsfonts}
\usepackage{bm}
\usepackage{algorithm}    
\usepackage{algpseudocode}  
\usepackage{float}
\DeclareSymbolFont{usualmathcal}{OMS}{cmsy}{m}{n}
\DeclareSymbolFontAlphabet{\mathcal}{usualmathcal}

\fancypagestyle{SPstyle}{
\fancyhf{}
\lhead{\colorbox{scipostblue}{\bf \color{white} ~SciPost Physics Codebases }}
\rhead{{\bf \color{scipostdeepblue} ~Submission }}

\fancyfoot[C]{\textbf{\thepage}}
}

\newcommand{\id}{\mathbb{1}}

\newcommand{\rr}{{\mathbf r}}

\newcommand{\tr}{{\mathrm{tr}}}
\newcommand{\supp}{{\mathrm{supp}}}

\renewcommand{\L}{{\mathcal L}}
\newcommand{\T}{{\mathcal T}}

\newcommand{\nmax}{{n_{\rm max}}}

\begin{document}

\pagestyle{SPstyle}

\begin{center}{\Large \textbf{\color{scipostdeepblue}{
\textsc{QCommute}: a tool for symbolic computation of nested commutators
in quantum many-body spin-1/2 systems 
}}}\end{center}

\begin{center}\textbf{
Oleg Lychkovskiy\textsuperscript{1,2},
Viacheslav Khrushchev\textsuperscript{3$$} and
Ilya Shirokov\textsuperscript{1,4$\star$}
}\end{center}

\begin{center}
{\bf 1} Skolkovo Institute of Science and Technology,\\
		Bolshoy Boulevard 30, bld. 1, Moscow 121205, Russia
\\
{\bf 2} Department of Mathematical Methods for Quantum Technologies,\\
		Steklov Mathematical Institute of Russian Academy of Sciences,\\
		8 Gubkina St., Moscow 119991, Russia
\\
{\bf 3} HSE University, 20 Myasnitskaya Ulitsa, Moscow, 101000, Russia
\\
{\bf 4} Moscow State University, Faculty of Physics, Moscow, 1199991 Russia
\\[\baselineskip]
$\star$ \href{mailto:email1}{\small shirokovie@my.msu.ru}
\end{center}

\section*{\color{scipostdeepblue}{Abstract}}
\textbf{\boldmath{%
We present \textsc{QCommute} \cite{QCommute}, a software tool implemented in C++ for symbolic computation of nested commutators between a Hamiltonian and local observables in quantum many-body spin-$1/2$ systems on one-, two-, and three-dimensional hypercubic lattices. The computation is performed algebraically directly in the thermodynamic limit, and the Hamiltonian parameters are kept symbolic. Importantly, this way the entire parameter space is covered in a single run. The implementation supports extensive parallelization to achieve high computational performance. \textsc{QCommute} can serve as a computational backend for Heisenberg-picture approaches to quantum dynamics in strongly correlated regimes, ranging from direct Taylor expansion in time to advanced techniques such as the recursion method.
}}

\vspace{\baselineskip}

\noindent\textcolor{white!90!black}{%
\fbox{\parbox{0.975\linewidth}{%
\textcolor{white!40!black}{\begin{tabular}{lr}%
  \begin{minipage}{0.6\textwidth}%
    {\small Copyright attribution to authors. \newline
    This work is a submission to SciPost Physics Codebases. \newline
    License information to appear upon publication. \newline
    Publication information to appear upon publication.}
  \end{minipage} & \begin{minipage}{0.4\textwidth}
    {\small Received Date \newline Accepted Date \newline Published Date}%
  \end{minipage}
\end{tabular}}
}}
}


\vspace{10pt}
\noindent\rule{\textwidth}{1pt}
\tableofcontents
\noindent\rule{\textwidth}{1pt}
\vspace{10pt}


\section{Introduction}
\label{sec:intro}

Quantitatively describing quantum many-body dynamics in nonperturbative regimes is, in general, a difficult task. While in one dimension a decisive progress has been achieved using matrix product state (MPS) techniques \cite{Schollwock_2011_Density-matrix,Fishman_2022_ITensor,Fishman_2022_Codebase}, in two and especially three dimensions the problem remains largely open. A variety of approaches have been developed to tackle this problem, including  variational and Monte-Carlo methods based on tensor network \cite{Dziarmaga_2022_Time,Hashizume_2022_Dynamical,DeNicola_2022_Entanglement,Kaneko_2023_Dynamics,Park_2025_Simulating,Pavevsic_2025_Constrained,mandra2025heuristic,Vovrosh_2026_Simulating} or neural network \cite{Schmitt_2020_Quantum,Vicentini_2022_NetKet,Vicentini_2022_Codebase,Sinibaldi_2026_Time-Dependent,chen2025convolutional,Vovrosh_2026_Simulating} representations of quantum many-body states, as well as  hybrid quantum-classical approaches \cite{Starkov_2018_Hybrid}. These methods are typically formulated in the Schr\"odinger picture of quantum mechanics. 

More recently, methods operating in the Heisenberg picture have attracted increasing attention \cite{Parker_2019,Khemani_2018_Operator,Rakovszky_2022_Dissipation-assisted,White_2023_Effective,Uskov_Lychkovskiy_2024_Quantum,ermakov2024unified,Loizeau_2025_Opening,Angrisani_2026_Simulating} and, in some cases, have demonstrated performance comparable to or exceeding that of more traditional Schr\"odinger-picture approaches \cite{Uskov_Lychkovskiy_2024_Quantum,Begusic_2024_Fast,Begusic_2025_Real-time,Teretenkov_2025_Pseudomode,shirokov2025quench,ermakov2025symbolic,Angrisani_2026_Simulating}. Most importantly, these include the recursion method \cite{Mori_1965_Continued-fraction,viswanath2008recursion,Parker_2019,Uskov_Lychkovskiy_2024_Quantum} and the sparse Pauli dynamics method \cite{Begusic_2024_Fast,Begusic_2025_Real-time}. The core computational primitive underlying all such Heisenberg methods is the evaluation of the commutator $[H,A]$ between a many-body Hamiltonian and an observable. The overall efficiency of Heisenberg methods hinges on the efficiency of the implementation of this core primitive.

Recently, the Julia package \textsc{PauliStrings.jl} has been released, providing an implementation of this primitive for spin-$1/2$ systems and building upon it to implement the recursion method~\cite{Loizeau_2025_Quantum,Loizeau_2025_Codebase,PauliStrings}. Another recent package with a related functionality, \textsc{PauliPropagation.jl} \cite{rudolph2025pauli}, places emphasis on the simulation of quantum circuits and sparse Pauli dynamics, with the additional capability to incorporate dissipation. A Python implementation used in the original sparse Pauli dynamics works~\cite{Begusic_2024_Fast,Begusic_2025_Real-time} is also publicly available~\cite{Begusic_GitHub}. Other groups working in the field have developed their own software~\cite{Roldan_1986_Dynamic,Florencio_1992_Quantum,Lindner_2010_Conductivity,Parker_2019,Huang_2019_Strange,De_2024_Stochastic,ermakov2025operator}  that has not been publicly released.

In this work, we present \textsc{QCommute} \cite{QCommute}, a tool for computing commutators in quantum translation-invariant spin-$1/2$ systems on hypercubic lattices. Written in C++, \textsc{QCommute} is designed for high computational efficiency, supports extensive parallelization, operates symbolically (retaining Hamiltonian parameters in analytic form), and works directly in the thermodynamic limit. We have recently employed \textsc{QCommute} to implement the recursion method for describing quench dynamics in one-, two-, and three-dimensional many-body systems \cite{shirokov2025quench}. Here, we describe the inner workings of the code and provide guidance for users. As an illustration, we present results on quantum many-body dynamics derived from the Taylor expansion of Heisenberg operators -- a conceptually simple approach that nevertheless yields remarkably accurate results and, in some cases, provides rigorous and extremely tight upper and lower bounds for dynamical quantities at early evolution times.

The rest of the paper is organized as follows. In the next section, we provide the necessary physical background and introduce the required notation, including the basics of quantum dynamics in the Heisenberg picture, as well as properties of Pauli strings and translation-invariant operators constructed from them. In Section \ref{sec:algorithm and implementation}, we present the algorithm underlying \textsc{QCommute} and its C++ implementation, outline installation and usage, and discuss performance.  In Section \ref{sec:application}, we apply \textsc{QCommute} to benchmark the short-time dynamics of exemplary spin systems. In the final section, we summarize  main features and capabilities of the package and highlight promising directions for future developments.

\section{Physical background}


\subsection{Quantum evolution in the Heisenberg picture}

A Heisenberg operator $A_t$ of a quantum observable is obtained from the Schr\"odinger operator $A$ of the same observable according to 
\begin{equation}\label{Heisenberg operator}
A_t=e^{i t H} A \, e^{-i t H}=e^{i t \L}A.
\end{equation}
Here $H$ is the Hamiltonian and 
\begin{equation}
\L\equiv [H,\bullet]
\end{equation}
is a {\it superoperator} (i.e. a linear map in the space of operators) referred to as Liouvillian. Both $A$ and $A_t$, as well as $H$, are self-adjoint operators.

In the many-body setting, $A_t$ can not be, in general, found explicitly (except noninteracting and certain interacting integrable models \cite{Prosen_2000_Exact,Lychkovskiy_2021_Closed,Gamayun_2022_Out-of-equilibrium,Teretenkov_2024_Exact,Ermakov_2025_Polynomially,Penc_2026_Linear}). Rather, it is to be approximated. As a rule, building blocks of such approximations are nested commutators
\begin{equation}\label{nested commutators}
\L^n A=\underbrace{[H,[H,\dots,[H}_{n\text{ times}},A]\dots]],
\end{equation}
where $n$ is referred to as {\it nesting order}.
The most straightforward approximation to $A_t$ is obtained  by truncating the Taylor expansion 
\begin{equation}\label{Taylor}
 A_t =\sum_{n=0}^\infty \frac{(it)^n}{n!} \L^n A
\end{equation}
at some $n=\nmax$, which is the maximal nesting order we are able to explicitly compute. More generally, one has a vast freedom to organize nested commutators in various linear combinations to improve the convergence, according to the formula 
\begin{equation}\label{exp expansion}
A_t =\sum_{n=0}^\infty \varphi_n(t)\, P_n(\L) A,
\end{equation}
where $\{ P_n(z)\}$ is a sequence of orthogonal polynomials and $\{ \varphi_n(t)\}$ is a related sequence of functions \cite{Gamayun_2025_Exactly}. In particular, the recursion method can be viewed as a specific implementation of the expansion~\eqref{exp expansion}, based on a sequence of orthogonal polynomials tailored to the given $H$ and $A$~\cite{Gamayun_2025_Exactly}. An alternative Heisenberg-picture approach, sparse Pauli dynamics\cite{Begusic_2024_Fast,Begusic_2025_Real-time,Angrisani_2026_Simulating}, is based on the trotterization of the unitary evolution in Eq.~\eqref{Heisenberg operator} and, at its computational core, likewise relies on the evaluation of commutators.

If one is able to compute $A_t$, two physical objects  become immediately accessible. The first  is the time-dependent post-quench expectation value, i.e. the  evolving expectation value of the observable after initialization in a state $\rho_0$ (assuming $\rho_0$ is explicitly known):
\begin{equation}
\langle A_t \rangle=\tr\,\rho_0 A_t.
\end{equation}
The second  is the infinite-temperature autocorrelation function of the observable,
\begin{equation}
C(t)=\tr\,A \,A_t/\tr\, A^2.
\end{equation}
It is worth noting that finite-temperature correlation functions can also be accessed, albeit using a considerably more sophisticated technique ~\cite{Angelinos_2026_Temperature}.

In the present paper, we introduce a software tool \textsc{QCommute} that efficiently computes nested commutators~\eqref{nested commutators} for spin-$1/2$ systems on hypercubic lattices. As an illustration of its performance, we compute approximate post-quench expectation values $\langle A_t \rangle$ and autocorrelation functions $C(t)$ for representative models. To this end, we employ the straightforward Taylor expansion~\eqref{Taylor}, which requires essentially no additional theoretical machinery, in contrast to more sophisticated approaches such as the recursion method. Specifically, we compute
\begin{align}
\langle A_t \rangle & \simeq\sum_{n=0}^{\nmax}\frac{q_n}{n!}t^n,\label{Taylor quench}\\ 
q_n & \equiv \tr \big( \rho_0\, (i \L)^n A\Big) \label{quench coefficient}
\end{align}
and
\begin{align}
C(t) & \simeq\sum_{n=0}^{\nmax}(-1)^n\frac{t^{2n}}{(2n)!} \, \mu_{2n}, \label{Taylor autocorrelation function}\\
\mu_{2n}& \equiv\tr\left( (i\L)^n  A\right)^2/\tr\, A^2\geq 0. \label{moment}
\end{align}
We refer to numbers $q_n$ as quench coefficients. Numbers $\mu_{2n}$ are commonly known as moments. One obtains eq. \eqref{Taylor autocorrelation function}  from eq. \eqref{Taylor} with the help of the identity  $\tr(A\,\L B)=-\tr((\L A) B)$, which implies, in particular, that the odd-order terms in the expansion of $C(t)$ vanish. 

A truncated Taylor expansion of a bounded function of time typically provides an excellent approximation at sufficiently small times but breaks down at larger times. This is precisely the behavior that will be observed below when applying Eqs.~\eqref{Taylor quench} and~\eqref{Taylor autocorrelation function}.

In fact, in the case of the autocorrelation function, the truncated Taylor expansion \eqref{Taylor autocorrelation function} yields a stronger result. Since eq. \eqref{Taylor autocorrelation function} is an alternating series, truncation at even (odd) order gives an upper (lower) bound on the value of the function, provided that, starting from some $n<\nmax$, the absolute values of the series terms decrease monotonically.\footnote{We consider only values of $t$ within the radius of convergence of the Taylor expansion, which is expected to be finite for two- and three-dimensional systems in the thermodynamic limit~\cite{Parker_2019} and infinite for one-dimensional systems~\cite{Araki_1969_Gibbs,Parker_2019}.} This assumption is, in fact, well justified: it is consistent with the universal operator growth hypothesis~\cite{Parker_2019} and with all case studies considered here and elsewhere \cite{Uskov_Lychkovskiy_2024_Quantum,Teretenkov_2025_Pseudomode,shirokov2025quench,ermakov2025symbolic}. Under this assumption, we get inequalities
\begin{equation}\label{bounds}
 \sum_{n=0}^{n_{\mathrm{max}}^{\mathrm{odd}}}(-1)^n\frac{t^{2n}}{(2n)!} \, \mu_{2n} \leq C(t)\leq \sum_{n=0}^{n_{\mathrm{max}}^{\mathrm{even}}}(-1)^n\frac{t^{2n}}{(2n)!} \, \mu_{2n},
\end{equation}
where $n_{\mathrm{max}}^{\mathrm{even}}$ and $n_{\mathrm{max}}^{\mathrm{odd}}$ are maximal available even and odd nesting orders, respectively. We will see that these inequalities provide very tight bounds useful for benchmarking other, more sophisticated but less controllable methods.

\subsection{Pauli strings}

We consider systems of spins $1/2$ on a hypercubic lattice of dimension $D\in \{1,2,3\}$. The lattice is infinite, i.e. no boundaries are assumed.  Sites of the lattice are labeled by a  $D$-dimensional integer vector $\rr \in \mathbb{Z}^D$. The site with all components of $\rr$ equal to 1 is called the {\it anchor site}. For example, the anchor site for the 3D lattice is labeled by $\rr=(1,1,1)$.

Operators are expanded in the basis of {\it Pauli strings}, which are  products of Pauli operators on different sites. A Pauli string  $P_\nu$ reads
\begin{equation}\label{Pauli string}
P_\nu=\prod_{k=1}^{w} \sigma^{\alpha_k}_{\rr_k}.
\end{equation}
Here $\sigma^{\alpha_k}_{\rr_k}$ with $\alpha_k\in\{x,y,z\}$ is a Pauli matrix on the site $\rr_k$, and the label $\nu$ is a shorthand for the multi-index
\begin{equation}\label{multi-index}
\nu=\big((\rr_1,\alpha_1), (\rr_2,\alpha_2),\dots,(\rr_w,\alpha_w)\big).
\end{equation}
The set $\supp(P_\nu)$ of site labels alone, 
\begin{equation}\label{support}
\supp(P_\nu)=(\rr_1,\rr_2,\dots,\rr_w),
\end{equation}
is refereed to as the {\it support} of the Pauli string \eqref{Pauli string}.  The identity operator is also regarded to be a Pauli string.

The site labels $\rr_k$ in eqs.~\eqref{multi-index},\eqref{support} are ordered lexicographically. We refer to the site $\rr_1$ in eqs.~\eqref{multi-index},\eqref{support} as the first site of the Pauli string. A Pauli string $\check P_\nu$ is {\it anchored} if its first site is the anchor site (e.g. $\rr_1 = (1,1,1)$ in the 3D case); the check mark on top of $\check P_\nu$ distinguishes anchored Pauli strings $\check P_\nu$ from general Pauli strings $P_\nu$.

Two important properties of a Pauli string are its {\it weight} and {\it size}. The weight $w(P_\nu)$ is the number of Pauli matrices in the Pauli string $P_\nu$. The size $s(P_\nu)$ is the length of the edge of the smallest hypercube that contains the support $\supp(P_\nu)$  of the Pauli string.

A product of two Pauli strings is also a Pauli string, perhaps multiplied by $(-1)$ or $(\pm i)$. This product is computed site-wise using identities
\begin{equation}\label{product of Pauli strings}
\sigma^\alpha_\rr    \sigma^\beta_\rr= \delta_{\alpha\beta}\id_\rr+i \sum_\gamma\varepsilon_{\alpha\beta\gamma} \sigma^\gamma_\rr,\qquad \alpha,\beta,\gamma\in{x,y,z}.
\end{equation}

The commutator $[P_\nu,P_{\nu'}]$ of two Pauli strings is either zero or again a Pauli string, up to a multiplicative numerical coefficient. If the supports of the two Pauli strings do not overlap, their commutator vanishes. The weight and the size of the commutator is limited by the respective weights and sizes of the Pauli strings as follows:
\begin{align}
w\big([P_\nu,P_{\nu'}]\big)&\leq w(P_\nu)+w(P_{\nu'})-1 \label{weight growth},\\
s\big([P_\nu,P_{\nu'}]\big)&\leq s(P_\nu)+s(P_{\nu'})-1 \label{size growth}.
\end{align}


\subsection{Translation invariance \label{subsec:translation}}

We address only translation-invariant Hamiltonians and observables. As an example, consider the transverse-field Ising Hamiltonian,
\begin{equation}\label{H Ising}
H=\sum_{\langle \rr \, \rr' \rangle} \sigma^x_{\rr}\sigma^x_{\rr'}+h_z \sum_{\rr} \sigma^z_{\rr},
\end{equation}
and the operator of the per-site magnetization along the $z$ direction,
\begin{equation}\label{magnetization}
A=N^{-1}\,\sum_{\rr} \sigma^z_{\rr}.
\end{equation}
The sum in eq. \eqref{magnetization} and the second sum in eq.\eqref{H Ising} run over all lattice sites, while the first sum in eq.\eqref{H Ising} runs over the pairs $\langle \rr \, \rr' \rangle$ of neighboring sites (i.e. over all edges) of the hypercubic lattice. The normalization factor $N^{-1}$ ($N$ being the total number of lattice sites) ensures that the observable is intensive (does not scale in the thermodynamic limit). This factor is purely formal: it  cancels in all end results, see eqs. \eqref{moments from c} and   \eqref{quench coefficients from c} below. 

To represent translation-invariant operators compactly, we introduce a linear superoperator $\T$ that distributes a local operator over the hypercubic lattice. Specifically, $\T$  maps a Pauli string $P_\nu$ to the sum of all strings obtained from $P_\nu$ by all possible  translations in all $D$ spatial directions. For example, for $D=3$, the Ising Hamiltonian~\eqref{H Ising} and the magnetization~\eqref{magnetization} can be written as
\begin{equation}\label{H Ising in T-form}
H=\T\left(  \sigma^x_{(1,1,1)}\sigma^x_{(2,1,1)}+ \sigma^x_{(1,1,1)}\sigma^x_{(1,2,1)}+ \sigma^x_{(1,1,1)}\sigma^x_{(1,1,2)}+h_z \sigma^z_{(1,1,1)}\right),
\end{equation}
and
\begin{equation}\label{magnetization in T-form}
A=N^{-1}\,\T \sigma^z_{(1,1,1)},
\end{equation}
respectively. Here site labels are  shown as explicit $D$-dimensional vectors. 

We say that a translation-invariant operator $A$ is in the {\it canonical form} if it can be represented as 
\begin{equation}\label{canonical form}
A=\T a
\end{equation}
where the {\it density} $a$ is a linear combination of anchored Pauli strings,
\begin{equation}\label{density}
    a=  \sum_\nu c_\nu \check P_\nu.
\end{equation}
We always assume that this sum contains a finite number of terms. 
Operators $H$ and $A$ in Eqs.~\eqref{H Ising in T-form} and \eqref{magnetization in T-form}, respectively, are written in the canonical form.

The key property of $\T$ is that it commutes with $\L$ whenever the corresponding Hamiltonian is translation-invariant. This allows one to simplify the computation of $\L A$ when $A$ is also translation-invariant:
\begin{equation}\label{actual computation sequence}
    \L A=N^{-1}\,\L \, \T a= N^{-1}\,\T \L a.
\end{equation}
This way, we first compute the commutator between $H$ and the density $a$ of the observable $A$ (remind that $a$ is a finite sum of Pauli strings), and then distribute the result over the whole lattice. Note that at the latter step every two Pauli strings that can be obtained from one another by some translation collapse to a single term, e.g.
\begin{equation}\label{canonicalization example}
 \T (\sigma^y_1\sigma^x_2+\sigma^y_2\sigma^x_3)=   \T (\sigma^y_1\sigma^x_2)+\T(\sigma^y_2\sigma^x_3)=  \T (\sigma^y_1\sigma^x_2)+\T(\sigma^y_1\sigma^x_2)=  \T ( 2\,\sigma^y_1\sigma^x_2),
\end{equation}
where $D=1$ is assumed. In the software implementation, this consequence of translation invariance leads to considerable memory savings.

Assume one has computed the $n$'th nested commutator and represented it in the canonical form,
\begin{equation}
    \L^n A=N^{-1}\,\T \sum_\nu c_\nu^{(n)} \check P_\nu,
\end{equation}
where the $ c_\nu^{(n)}$ are numerical coefficients that are real (imaginary) for even (odd) $n$.  Then one immediately obtains  the $(2n)$'th moment \eqref{moment}:
\begin{equation}\label{moments from c}
\mu_{2n}=\sum_\nu | c_\nu^{(n)}|^2/\sum_\nu | c_\nu^{(0)}|^2.
\end{equation}
Here $c_\nu^{(0)}$ are expansion coefficients of the density $a$ of the observable $A$: $a= \sum_\nu c_\nu^{(0)} \check P_\nu$.

To obtain quench coefficients \eqref{quench coefficient}, an initial state should be specified. We focus on translation-invariant product states 
\begin{equation}\label{product state}
\rho_0=
\bigotimes_{\rr}
\frac12
\left(
\id_\rr + \bm{p}\bm{\sigma}_\rr
\right),
\end{equation}
where $\bm{p}\bm{\sigma}_\rr\equiv p_x\sigma^x_\rr+p_y\sigma^y_\rr+p_z\sigma^z_\rr$ and $\bm{p}$ is a real three-dimensional {\it polarization vector}  satisfying $|\bm{p}| \le 1$.

For this initial state the $n$'th quench coefficient \eqref{quench coefficient} reads
\begin{equation}\label{quench coefficients from c}
    q_n=\sum_\nu c_\nu^{(n)} \, \Upsilon(\check P_\nu).
\end{equation}
Here $\Upsilon( P_\nu)$ is a real number obtained from the Pauli string $ P_\nu$ by substituting each Pauli matrix by a corresponding component of the polarization vector, $\sigma^\alpha_\rr\rightarrow p_\alpha$, e.g.
\begin{equation}\label{Upsilon}
\Upsilon(\sigma^y_1\sigma^x_5\sigma^y_6\sigma^z_8)=p_x \, p_y^2 \, p_z.
\end{equation}

The notions of weight and size of a Pauli string can be extended to an arbitrary operator. To this end, one expands the operator in the basis of Pauli strings and defines its weight (size) as the maximum of the weights (sizes) of the Pauli strings appearing in the expansion with nonzero coefficients. For example, $w(H)=s(H)=2$ for the Ising Hamiltonian~\eqref{H Ising}.

It is easy to see that inequalities \eqref{weight growth},\eqref{size growth} hold for arbitrary operators. In particular, inequality \eqref{size growth} implies that the action of the Liouvillian increases the size of an operator by at most $(s(H)-1)$.  One consequence is that  the computation of $\L^n A$ yields essentially identical results for a system on an infinite lattice and for a finite system with periodic boundary conditions, provided that its linear size exceeds $s(A)+n(s(H)-1)$. This observation should be kept in mind when comparing results for infinite and finite systems.

\section{Algorithm and implementation \label{sec:algorithm and implementation}}

\subsection{Algorithm}\label{sec:algorithm}

The algorithm iteratively computes the sequence of nested commutators $A^{(n)}$ defined by
\begin{align}
    A^{(0)} &= A  \qquad  (\text{initial observable}), \\
    A^{(n+1)} &=\L A^{(n)}\equiv [H, A^{(n)}].
\end{align}
As long as the observable $A$ and the Hamiltonian $H$ are translation-invariant, they are determined by their densities $h$ and $a$, respectively,\footnote{The density $h$ should not be confused with the magnetic field constant $h_z$ entering the Hamiltonian \eqref{H Ising}.} 
\begin{equation}
H=\T h,\qquad A=N^{-1}\,\T a,
\end{equation}
as explained in the previous Section.  Nested commutators are also translation-invariant and, therefore, are also determined by their densities. The computation of these densities is simplified by the fact that the Liouvillian and the translation superoperator commute, as described in Eq.~\eqref{actual computation sequence}. Thus, at the highest level, the algorithm is as follows:

\begin{algorithm}[H]
\caption{Computation of nested commutators}
\label{alg:nested}
\begin{algorithmic}[1]
\State \textbf{Input:} density of Hamiltonian $h$, density of observable $a$, maximal nesting order $\nmax$
\State \textbf{Output:} sequence $\{a^{(n)}\}_{n=0}^\nmax$ of densities of nested commutators $\L^nA$
\State $ a^{(0)} \gets a$
\For{$n \gets 1$ to $\nmax$}
    \State $a^{(n)} \gets \text{Commute}(\T h, a^{(n-1)})$
\EndFor    
\State \Return $\{a^{(n)}\}_{n=0}^\nmax$
\end{algorithmic}
\end{algorithm}

The central routine of this algorithm is \texttt{Commute}. It performs the following procedure:

\begin{algorithm}[H]
\caption{\texttt{Commute}$(\T h, a)$ routine}
\label{alg:nested}
\begin{algorithmic}[1]
\State \textbf{Input:}  density of Hamiltonian $h=\sum_{\nu\in \Omega_H } c_\nu^H \,\check P_\nu$, density of observable $a= \sum_{\nu\in \Omega_A } c_\nu^A\, \check P_\nu$
\State \textbf{Output:} density of the commutator $[H, A]$   
\State $res \gets 0$
\For{each $\nu' \in \Omega_A$}
    \For{each $P_\nu$ from $H=\T h$ such that $\supp(\check P_{\nu'}) \cap \supp(P_\nu) \neq \varnothing$}
        \State $comm \gets c_\nu^H \, c_{\nu'}^A\,[P_\nu, P_{\nu'}]$ 
        \State $res \gets \mathrm{Add}( comm,\, res)$
        \State $res \gets \mathrm{Canonicalize}( res)$
        \State $res \gets \mathrm{Simplify}( res)$
    \EndFor
\EndFor
\State \Return $res$
\end{algorithmic}
\end{algorithm}
\noindent
Here, in line 1, sets $\Omega_H$ and $\Omega_A$ contain labels of anchored Pauli strings that enter $h$ and $a$,  respectively. In line 5, one takes every $P_\nu$ (not necessarily anchored) that enters $H$ and has a nonzero overlap with an anchored Pauli string $\check P_{\nu'}$ from $a$. In line 6, if  $P_\nu$ is not anchored,  $c_\nu^H$ is the coefficient associated to the anchored Pauli string $\check P_\nu$ related to  $P_\nu$ by a translation. In line 8, the \texttt{Canonicalize} command brings the operator to the canonical form, replacing each non-anchored Pauli  string $P_\nu$ by its anchored counterpart $\check P_\nu$  related to $P_\nu$  by a translation (see eq.\eqref{canonicalization example} for an example).  In line 9, the \texttt{Simplify} command  collects like terms and removes those with zero coefficients  (see eq.\eqref{canonicalization example}).

Importantly, all coefficients multiplying Pauli strings in the expansion of $\L^n A$ in the Pauli-string basis are treated symbolically. Specifically, the coefficients of Pauli strings in the Hamiltonian and the observable are restricted to be be either integers or symbolic parameters, as in Eqs.~\eqref{H Ising in T-form} and~\eqref{magnetization in T-form}. The \texttt{Commute} routine is carried out algebraically, so that all coefficients of Pauli strings in $\L^n A$ become polynomials in these parameters with integer coefficients. In what follows, we refer to these polynomials as {\it coefficient polynomials} for brevity.

\subsection{Implementation}\label{sec:implementation}


The algorithmic framework described in Section~\ref{sec:algorithm} is implemented in C++17, with a strong focus on high performance and the exactness of symbolic computations. Below we highlight the key architectural features of the \textsc{QCommute} package:

\begin{itemize}
    \item \textbf{Data structures:} The physical operators are mapped directly to efficient C++ structures. The spatial coordinate of a lattice site, represented mathematically by a $D$-dimensional integer vector $\mathbf{r}$, is implemented as a template class \texttt{Point<D>}. A local Pauli operator $\hat{\sigma}^\alpha_{\mathbf{r}}$ is represented by the \texttt{Sigma} structure, which encapsulates the \texttt{Point<D>} coordinate and an integer indicating the Pauli matrix type $\alpha\in\{x,y,z\}\cong\{1,2,3\}$.  Coefficient polynomials are stored in a dedicated \texttt{Poly} class that contains two vectors. The first vector encodes monomial exponents as 8-bit integers, with individual monomials separated by the delimiter value 254 (consequently, the maximum allowed exponent of each variable is 253). The second vector stores the corresponding coefficients as arbitrary-precision integers, as described below. 
    
    \item \textbf{Arbitrary-precision integer arithmetic:} Computing nested commutators generates a combinatorially large number of terms, and the coefficient polynomials (in particular, integer coefficients of these polynomials) grow extremely fast. Standard 64-bit integer types would inevitably overflow for large nesting orders. To prevent this, the code utilizes the GMP library (\texttt{gmpxx}) for arbitrary-precision integer arithmetic, ensuring mathematically exact integer coefficients regardless of the nesting order.
    
    \item \textbf{Disk swapping and chunking:} The core challenge of computing nested commutators is the exponential growth of the result with nesting order, which rapidly exhausts available RAM. To circumvent this, \textsc{QCommute} does not keep the entire commutator expression in memory. Instead, it dynamically splits the data into manageable chunks and writes intermediate expressions to the disk as binary \texttt{.comm} files. These files are stored in dedicated subdirectories, one for each nesting order. This way, the program is able to compute deeply nested commutators even on personal computers.
\end{itemize}

\subsection{Installing and running \textsc{QCommute}}

\label{sec:installation}

The \textsc{QCommute} package is distributed as open-source software via GitHub~\cite{QCommute}. The core computations rely on exact symbolic algebra and arbitrary-precision integer arithmetic. Thus, the only strict external dependency is the GNU Multiple Precision Arithmetic Library (GMP/gmpxx). A C++17 compliant compiler, such as GCC or Clang, is required.

\subsubsection*{Compilation}
For UNIX-like environments (Linux, macOS), the recommended approach for compiling the software is utilizing the \textit{Meson} build system:
\begin{verbatim}
git clone https://gitlab.com/viacheslav_hrushev/commutators
cd commutators
meson setup --buildtype=release build-release
cd build-release
meson compile
\end{verbatim}
This sequence configures the project with high optimization flags and produces the executable binary named \texttt{main}.

Alternatively, users can compile the program manually directly via GCC. This is particularly useful for Windows users or those utilizing the provided \texttt{tasks.json} file in Visual Studio Code. The equivalent manual compilation command reads
\begin{verbatim}
g++ *.cpp -o main -std=c++17 -O3 -lgmpxx -lgmp
\end{verbatim}

\subsubsection*{Model configuration file}

The model, i.e. the Hamiltonian, the observable and the maximal nesting order, is specified via a plain-text configuration file (e.g., \texttt{Comm.txt}). The program parses this file to construct symbolic representations of operators. The file consists of three main sections, denoted by \texttt{H:}, \texttt{O:}, and \texttt{n:}.

For example, a configuration file for the Ising Hamiltonian \eqref{H Ising in T-form} on the cubic lattice, observable \eqref{magnetization in T-form} and maximal nesting order $\nmax=12$ reads
\begin{verbatim}
H:
[1]*Sx(1,1,1)*Sx(2,1,1)
[1]*Sx(1,1,1)*Sx(1,2,1)
[1]*Sx(1,1,1)*Sx(1,1,2)
[1,1]*Sz(1,1,1)
O:
[1]*Sz(1,1,1)
n:
12
\end{verbatim}

The details of the syntax are as follows. 

\begin{itemize}
    \item \textbf{Pauli operators and dimensionality:} \texttt{Sx($r_x,r_y,r_z$)}, \texttt{Sy($r_x,r_y,r_z$)}, \texttt{Sz($r_x,r_y,r_z$)} stay for $\sigma^x_\rr$, $\sigma^y_\rr$, $\sigma^z_\rr$, respectively, with $\rr=(r_x,r_y,r_z)$.   The spatial dimension $D$ of the model is implicitly defined by the number of components of $\rr$  (e.g. \texttt{Sx(1)} is used in the  1D case, \texttt{Sx(1,1)} -- in the 2D case, and \texttt{Sz(1,1,1)} -- in the 3D case). 
    \item \textbf{Hamiltonian (\texttt{H:}):} The density of the model Hamiltonian is specified as a finite sum of terms, each term being a product of a symbolic coefficient (see below) and an anchored Pauli string.
    \item \textbf{Observable (\texttt{O:}):} The density of the observable is specified analogously to the density of the Hamiltonian.  The formal normalization factor $N^{-1}$ is omitted.
    \item \textbf{Nesting order (\texttt{n:}):} A positive integer specifying the maximum nesting order $\nmax$ to be computed.
\end{itemize}

\subsubsection*{Coefficient polynomial syntax}
A key feature of \textsc{QCommute} is that coupling constants are treated symbolically as polynomials of abstract system parameters (e.g., $h_1, h_2, \dots$) with integer coefficients. These coefficient polynomials are encoded in square brackets \texttt{[...]}. Terms of the polynomial are separated by a semicolon (\texttt{;}). Each term is a comma-separated list where the first number is the integer coefficient, and the subsequent numbers are the powers of the abstract parameters $h_i$. For example:
\begin{itemize}
    \item \texttt{[1]} stands for $1$.
    \item \texttt{[2,2]} stands for $2 h_1^2$.
    \item \texttt{[1;1,1,1;1,0,1;2,2,1]} stands for $1 + h_1 h_2 + h_2 + 2 h_1^2 h_2$.
\end{itemize}

\subsubsection*{Execution Pipeline}
The execution of the \textsc{QCommute} package generally follows a two-step pipeline: computing nested commutators up to the given nesting order $\nmax$ and subsequently extracting moments \eqref{moment} or quench coefficients \eqref{quench coefficient} form the nested commutators.

\subsubsection*{Step 1: Computing nested commutators}
By default, the program writes the results to a \texttt{./tmp} directory, but a custom output directory can be specified using the \texttt{--res-dir} flag:
\begin{verbatim}
./main Comm.txt --res-dir ./results_dir
\end{verbatim}
To manage memory efficiently, the program does not keep all commutators in RAM. Instead, it creates a subdirectory for each nesting order (e.g., \texttt{0}, \texttt{1}, \dots, \texttt{n}) containing binary \texttt{.comm} files with the computed Pauli strings. 

\subsubsection*{Step 2: Computing moments and quench coefficients}
Once the commutators are generated, the user can invoke specific sub-commands pointing to the generated directory to extract moments \eqref{moments from c} or quench coefficients \eqref{quench coefficients from c}. 

To compute moments \eqref{moments from c}, the \texttt{momentum} sub-command is invoked:
\begin{verbatim}
./main momentum ./results_dir > momentums.txt
\end{verbatim}
The resulting moments are (in general, multivariate) polynomials of the model parameters with integer coefficients. The output file is produced in \textsc{Wolfram Mathematica}\textsuperscript{\textcopyright}-readable form, with the Hamiltonian parameters denoted by $h0,h1,\dots$. For example, for the 1D mixed-field Ising model, whose Hamiltonian \eqref{H 1D ising}  contains two parameters (see below), the magnetization \eqref{magnetization} as the observable and the nesting order $\nmax=3$, the output file reads 
\begin{verbatim}
moments = {
8 + 4 h0^2,
128 + 128 h1^2 + 192 h0^2 + 16 h0^2 h1^2 + 16 h0^4,
2048 + 6144 h1^2 + 2048 h1^4 + 7680 h0^2 + 5888 h0^2 h1^2 + 64 h0^2
h1^4 + 1920 h0^4 + 128 h0^4 h1^2 + 64 h0^6
};
\end{verbatim}
To compute quench coefficients \eqref{quench coefficients from c}, the \texttt{quench} sub-command is used:
\begin{verbatim}
./main quench ./results_dir > quenches.txt
\end{verbatim}
The resulting quench coefficients are multivariate polynomials of the model parameters  and components of the polarization vector $\bf p$  with integer coefficients. For the same exemplary model as above, the output file reads 
\begin{verbatim}
quenchCoef = {
py (-2 h0) + px py (-4),
pz (8 + 4 h0^2) + py^2 (8 h1) + px (-4 h0 h1) + px pz (16 h0) + px^2 
(-8 h1) + px^2 pz (8),
py (-48 h0 + -8 h0 h1^2 + -8 h0^3) + py pz (64 h0 h1) + px py (-64 + 
-64 h1^2 + -48 h0^2) + px py pz (64 h1) + px^2 py (-48 h0)
};
\end{verbatim}
Note that the imaginary unit is omitted  in this file, so to get a quench coefficient \eqref{quench coefficients from c} of the odd order one should multiply the entry of this file by $i$.

In addition, the program accepts several auxiliary flags to control its behavior and manage system resources:
\begin{itemize}
    \item \texttt{--memory-limit ML}: Sets the memory limit for the program to \texttt{ML} megabytes. Because the number of Pauli strings grows exponentially, if the memory usage exceeds the RAM size, the OS automatically swaps the data to the disk. However, this swap procedure is hardware-dependent and can result in severe performance degradation or out-of-memory errors. Users are highly recommended to explicitly enforce a memory limit below the available RAM (with a reservation for other processes running simultaneously and consuming a part of RAM) by specifying this flag. The program will then safely handle the data chunking and its own optimized disk-writing.
    
    \item \texttt{--res-dir DIRECTORY}: Specifies a custom path where the directories with the results are to be saved (the default is \texttt{./tmp}).
    
    \item \texttt{--force}: Automatically deletes the contents of the target result directory before starting the computation. If the directory is not empty and this flag is omitted, the program will stop with an error message to prevent accidental mixing of old and new data.
    
    \item \texttt{--text-output}: Prints the final computed commutator to the standard output in a human-readable format. While useful for debugging at small $n$, it is not recommended for large nesting order due to a significant input/output overhead. Here is an example of such output for the 3rd commutator of the 1D Ising Model \eqref{H 1D ising}:
    \begin{verbatim}
i*[-48,1,0;-8,1,2;-8,3,0]*Sy(1)
i*[32,0,1]*Sx(3)*Sy(1)*Sz(2)
i*[32,0,1]*Sx(1)*Sy(3)*Sz(2)
i*[-48,1,0]*Sx(1)*Sx(3)*Sy(2)
i*[32,1,1]*Sy(2)*Sz(1)
i*[-32;-32,0,2;-24,2,0]*Sx(2)*Sy(1)
i*[32,1,1]*Sy(1)*Sz(2)
i*[-32;-32,0,2;-24,2,0]*Sx(1)*Sy(2)
\end{verbatim}
\end{itemize}

Finally, the package includes a standalone \texttt{validate [FILE1] [FILE2]} sub-command. This utility accepts two human-readable files containing sums of Pauli strings and explicitly reports whether the sums match, making it a valuable tool for debugging and cross-verifying results with other software.

\begin{table}[t]
\centering
\caption{Maximal computed nesting order $\nmax$ for the Ising model \eqref{H Ising} (2D and 3D cases) or \eqref{H 1D ising} (1D case) reported in various studies. \label{table}}
\label{tab:bn_comparison}
\begin{tabular}{lccccc}
\hline
   &2019 &  2021 &  2024 & 2024 & 2025\\
   & Parker {\it et al.} & Noh & De {\it et al.} & Uskov \& Lychkovskiy & \textsc{QCommute}  \\
        & \cite{Parker_2019} & \cite{Noh_2021}&  \cite{De_2024_Stochastic}&\cite{Uskov_Lychkovskiy_2024_Quantum}& \cite{shirokov2025quench} \\
\hline
1D &   30 & 38 & 35 & 45 & 48 \\
2D &   -- & -- & 16 & 21 & 23 \\
3D &   -- & -- & -- & -- & 17 \\
\hline
\end{tabular}
\end{table}

\subsection{Crosschecks}
We have validated the results produced by our package by comparing them with results obtained for lower nesting orders using three independent software tools. First, for one-dimensional systems, we implemented a simplified version of our algorithm in \textsc{Wolfram Mathematica}\textsuperscript{\textcopyright}. The second tool is a computer program (currently unsupported) written by Filipp Uskov in the \textsc{FORM} language. This program was used in Refs.~\cite{Uskov_Lychkovskiy_2024_Quantum,Teretenkov_2024_Exact}, where results for two-dimensional systems were reported. Finally, for three-dimensional systems, we compared our results with those obtained using the code by Igor Ermakov \cite{ermakov2025operator,ermakov2025symbolic}. In addition, we have cross-checked our results against exactly solvable cases -- the one-dimensional transverse-field Ising model, as well as the classical Ising model in arbitrary dimension, given by eq.~\eqref{H Ising} with $h_z=0$.

\subsection{Performance}

\begin{figure}[t]
    \centering
    \includegraphics[width=\textwidth]{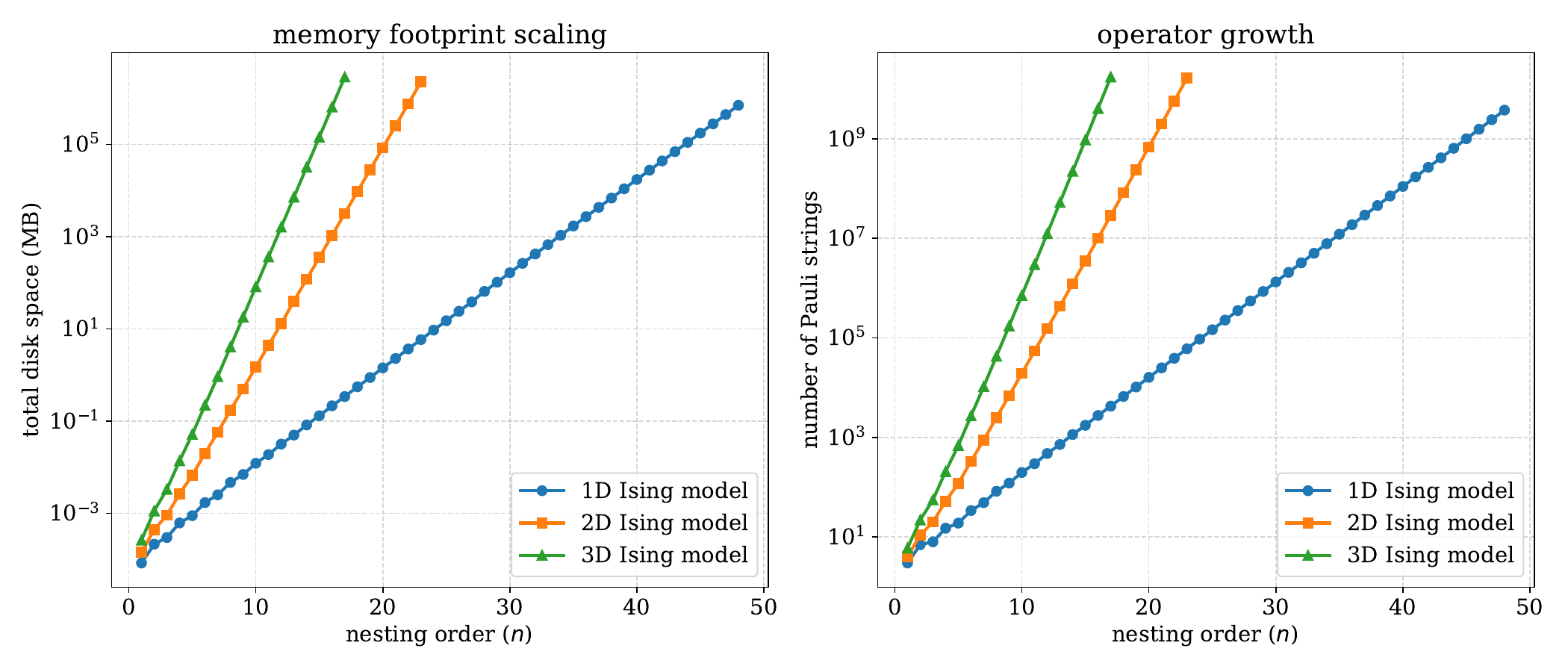}
    \caption{Scaling of the memory footprint (left) and the total number of unique Pauli strings (right) as a function of the nesting order $n$ for the 2D and 3D transverse-field Ising model \eqref{H Ising}, and the 1D mixed-field Ising model \eqref{H 1D ising}. The algorithm avoids out-of-memory errors by dynamically caching operators on the disk.}
    \label{fig:performance}
\end{figure}

\begin{figure}[t]
    \centering
    \includegraphics[width=\textwidth]{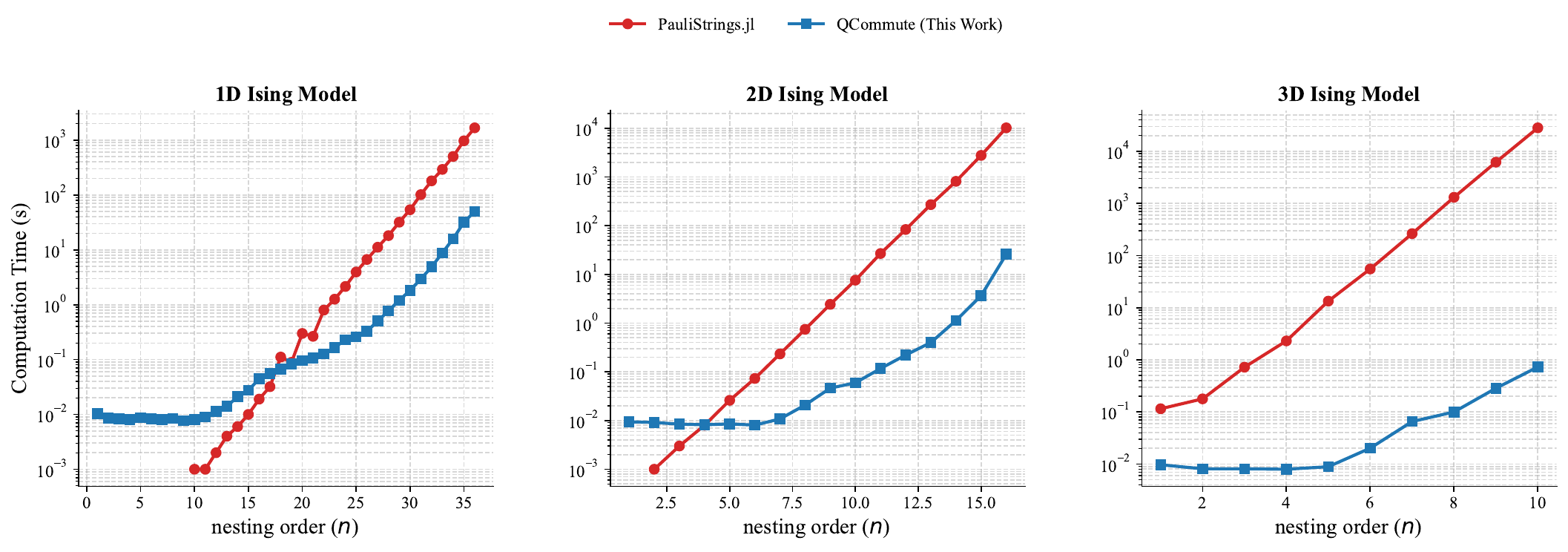}
    \caption{Performance comparison between \textsc{QCommute} and \textsc{PauliStrings.jl}~\cite{Loizeau_2025_Quantum,Loizeau_2025_Codebase,PauliStrings} for computing nested commutators in the 1D mixed-field Ising model \eqref{H 1D ising} and the 2D and 3D transverse-field Ising models \eqref{H Ising}.  The computational time (in seconds) is plotted on a logarithmic scale against the nesting order $n$. Missing data points for the \textsc{PauliStrings.jl}  at low nesting orders in the one- and two-dimensional case indicate that the execution times are negligible and fall below the precision limit of the system timer. The software has been run on the server containing four Intel Xeon Gold 6140 processors operating at 2.30 GHz and 512 GB of RAM. }
    \label{fig:benchmarks}
\end{figure}

In order to highlight the growth of operator size and computational complexity with nesting order and demonstrate the efficiency of \textsc{QCommute} in addressing this growth,  we perform computations for the Ising model on 1D, 2D, and 3D hypercubic lattices. The observable considered is the magnetization~\eqref{magnetization}. In the 2D and 3D cases, the Hamiltonian is given by eq. \eqref{H Ising}. In one dimension, the Hamiltonian \eqref{H Ising} is exactly solvable; as a consequence, $\L^n A$ do not feature exponential growth and admit a relatively simple analytical form \cite{Prosen_2000_Exact,Lychkovskiy_2021_Closed,Ermakov_2025_Polynomially}. For this reason, in the 1D case we address the nonintegrable {\it mixed-field} Ising model with the  Hamiltonian
\begin{align}
    H_{1D} &= \sum_{j} \sigma^x_j \sigma^x_{j+1} + h_x \sum_{j} \sigma^x_j + h_z \sum_{j} \sigma^z_j. \label{H 1D ising} 
\end{align}

As shown in Fig.~\ref{fig:performance}, while the computational complexity grows exponentially at any dimensionality, the growth exponent is highly sensitive to the dimensionality. In the 1D case, this exponent is relatively low, allowing the program to comfortably reach $\nmax=47$ at a 1 TB RAM server without swapping. At this nesting order, the expansion produces approximately $2.4 \times 10^9$ unique Pauli strings, occupying merely 428 GB of disk space. 

Conversely, in the 2D and 3D cases the exponent is considerably larger, inducing a rapid exponential explosion of the number of Pauli strings and the memory usage. The true advantage of the disk-caching algorithm becomes evident already in the 2D case. Standard approaches relying solely on Random Access Memory (RAM) would face an inevitable out-of-memory (OOM) crash before reaching $n\simeq 20$ (at a 1 TB RAM server). By shifting the workload to NVMe (non-volatile memory express) storage, our software successfully handles over $1.7 \times 10^{10}$ unique terms at $n=23$ (consuming roughly 2.16 TB of disk space).

In Table~\ref{table}, we list state-of-the-art maximal nesting orders computed for the transverse Ising model~\eqref{H Ising} on two- and three-dimensional hypercubic lattices and for the mixed-field one-dimensional Ising model~\eqref{H 1D ising}, as reported in the literature, and compare them with the results obtained using \textsc{QCommute}~\cite{shirokov2025quench}. We include only results free of boundary effects, i.e., those obtained either directly in the thermodynamic limit or for sufficiently large finite systems such that, for a given $\nmax$, the results coincide with the thermodynamic-limit values. The comparison demonstrates that \textsc{QCommute} performs favorably against other available tools, consistently exceeding reported benchmarks.

To further highlight the efficiency of \textsc{QCommute}, we perform a direct performance comparison against \textsc{PauliStrings.jl}~\cite{Loizeau_2025_Quantum,Loizeau_2025_Codebase,PauliStrings} -- a leading alternative package that, among other features, computes nested commutators for many-body spin-$1/2$ systems. Both codes were run on an identical server running Ubuntu 24.04.4 LTS, equipped with four Intel Xeon Gold 6140 processors (2.30 GHz) and 512 GB of RAM, configured as sixteen 32 GB modules. 
The results of this comparison are shown in Fig.~\ref{fig:benchmarks}. As can be seen, \textsc{QCommute} exhibits substantially higher efficiency than \textsc{PauliStrings.jl}, particularly for large nesting orders, where memory and CPU requirements are most demanding. 

It is worth noting that ref.~\cite{Uskov_Lychkovskiy_2024_Quantum}, although sharing one author with the present work, reports results obtained using a different tool written in the \textsc{FORM} language. Among the listed implementations, only this code and \textsc{QCommute} support symbolic computations with Hamiltonian parameters kept analytic, whereas other tools perform computations for specific numerical values of these parameters. We also note that Ref.~\cite{De_2024_Stochastic} reports two-dimensional results for a slightly more complex mixed-field Ising model.

\section{An application: short-time dynamics of quantum Ising model \label{sec:application}}

\begin{figure}
 \centering
    \includegraphics[width=1\textwidth]{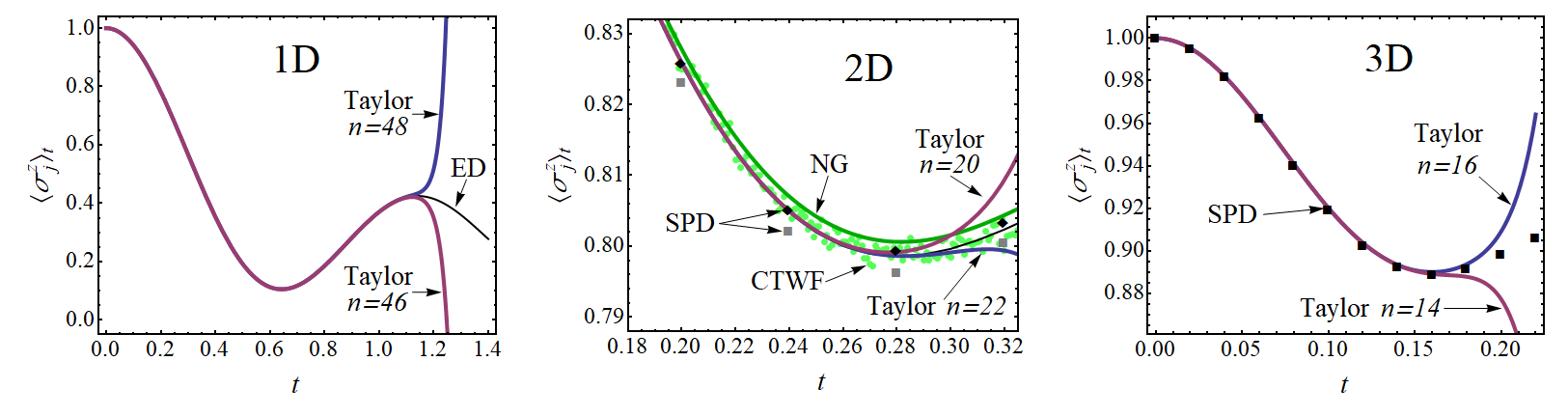}
    \caption{
    Quench dynamics of the quantum Ising model on one-, two-, and three-dimensional hypercubic lattices. The observable is the polarization per site in the $z$-direction, and the initial state is a product state fully polarized along $z$. The blue (magenta) line shows the Taylor expansion~\eqref{Taylor quench} truncated at the maximal available even order $n_{\rm max}^{\rm even}$ (at the order $(n_{\rm max}^{\rm even}-2)$), see text for details. \textbf{1D:} Results are shown for the mixed-field Ising model~\eqref{H 1D ising} with $h_x=h_z=1$. The black line represents the exact diagonalization result, which is essentially indistinguishable from the thermodynamic-limit result on the time scale shown. \textbf{2D:} Results are shown for the transverse-field Ising model~\eqref{H Ising} at the critical field $h_z=3.04438$. For comparison, we also show results obtained using other methods: black line -- iPEPS~\cite{Dziarmaga_2022_Time}, green line -- NG~\cite{Sinibaldi_2026_Time-Dependent}, green points -- CTWF~\cite{chen2025convolutional}, black diamonds -- SPD with Trotter step $\delta t=0.001$~\cite{Begusic_2025_Real-time}, and gray squares -- SPD with $\delta t=0.004$~\cite{Begusic_2025_Real-time}. \textbf{3D:} Results are shown for the transverse-field Ising model~\eqref{H Ising} at the critical field $h_z=5.15813$. The black line shows the SPD result~\cite{Begusic_2025_Real-time}.  
    }
    \label{fig:quench}
\end{figure}
Here we apply \textsc{QCommute} to study short-time dynamics of the one-dimensional mixed-field Ising model~\eqref{H 1D ising} and the two- and three-dimensional transverse-field Ising model~\eqref{H Ising}. As the observable, we consider the total magnetization~\eqref{magnetization}.


We first consider a quantum quench from the product state~\eqref{product state} polarized along the $z$-direction, ${\bf p}=(0,0,1)$, with the results shown in Fig.~\ref{fig:quench}. We compute the quench coefficients~\eqref{quench coefficient} and use them to approximate the post-quench dynamics of the magnetization via the truncated Taylor expansion~\eqref{Taylor quench}. For this particular initial state, the odd quench coefficients~\eqref{quench coefficient} vanish identically. This follows from T-invariance and can be verified directly by taking the complex conjugate of Eq.~\eqref{quench coefficient}. Accordingly, we present results for truncation orders $n_{\rm max}^{\rm{even}}$ and $(n_{\rm max}^{\rm{even}}-2)$. The coincidence of the two curves determines the time interval over which the approximation is reliable.

We compare our results with other available methods. In one dimension, we benchmark against exact diagonalization (ED) of a spin chain with $N=12$ sites and periodic boundary conditions. We have verified that this system size is sufficient for the results to be effectively indistinguishable from those in the thermodynamic limit on the time scales considered.

In two dimensions, we compare our results with those obtained using the infinite projected entangled pair state (iPEPS)~\cite{Dziarmaga_2022_Time}, the sparse Pauli dynamics (SPD)~\cite{Begusic_2025_Real-time}, the time-dependent neural Galerkin (NG)~\cite{Sinibaldi_2026_Time-Dependent} methods, and the time-dependent variational principle with convolutional transformer wave functions (CTWF)~\cite{chen2025convolutional}. As noted recently in Ref.~\cite{Park_2025_Simulating}, there is a small discrepancy between the iPEPS and SPD results on the one hand and the NG results on the other. Our results agree with the iPEPS and most accurate (i.e. those obtained with the smallest available Trotter time step) SPD data, thus providing an independent high-precision benchmark. In contrast, the CTWF results exhibit noise at a level comparable to this discrepancy and therefore do not resolve it. 

In three dimensions, we compare our results with those obtained using SPD, which, to the best of our knowledge, provides the only independent results currently available in three dimensions. We find excellent agreement.

\begin{figure}[t!]
    \centering
    \includegraphics[width=1\textwidth]{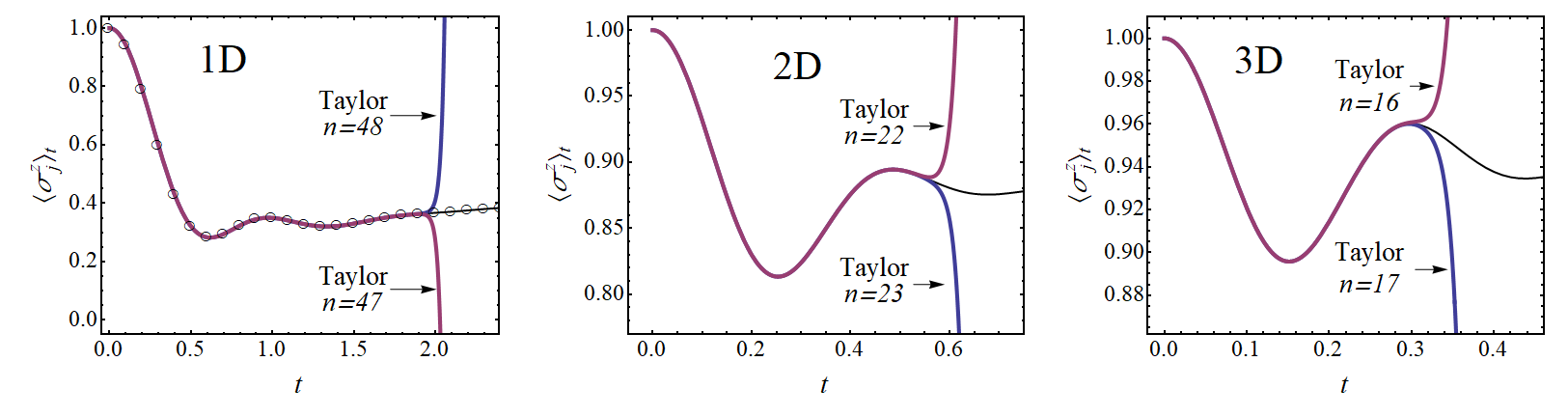}
    \caption{
    Autocorrelation function for the quantum Ising model on one-, two-, and three-dimensional hypercubic lattices. The observable and the Hamiltonians are the same as in Fig. \ref{fig:quench}. The blue (magenta) line shows the Taylor expansion~\eqref{Taylor autocorrelation function} truncated at the maximal available order $\nmax$ (at the order $(\nmax-1)$). The black line shows the results of the recursion method~\cite{Uskov_Lychkovskiy_2024_Quantum,shirokov2025quench}. Additionally, in the 1D case the exact diagonalization result is shown by open circles.
    }
    \label{fig:correlation function}
\end{figure}

Next, we consider the autocorrelation function~\eqref{Taylor autocorrelation function}. The upper and lower bounds~\eqref{bounds} obtained from the Taylor expansion are shown in Fig.~\ref{fig:correlation function}. In one dimension, we again benchmark against essentially exact results from exact diagonalization. In two and three dimensions, where independent benchmarks are scarce, we compare instead with results obtained using the recursion method~\cite{Uskov_Lychkovskiy_2024_Quantum,shirokov2025quench}, which relies on the same moments $\mu_{2n}$ that enter Eqs.~\eqref{Taylor autocorrelation function},\eqref{bounds}. One can see from Fig.~\ref{fig:correlation function} that the recursion-method results lie within the Taylor bounds~\eqref{bounds}, as they should. Observe that, up to a certain time, the two-sided bound~\eqref{bounds} is extremely tight, effectively yielding the exact autocorrelation function. 

\section{Summary and outlook}

To summarize, the \textsc{QCommute} package \cite{QCommute} is a highly efficient tool for computing nested commutators between  translation-invariant operators in spin-$1/2$ systems on one-, two-, and three-dimensional hypercubic lattices. It has the following distinctive features:
\begin{itemize}
\item it operates directly in the thermodynamic limit (infinite lattice, no boundary effects);
\item it treats Hamiltonian parameters symbolically, so that a single run produces results across the entire parameter space;
\item it employs extensive and highly optimized parallelization.
\end{itemize}

This combination of features makes \textsc{QCommute} unique among existing tools for computing nested commutators, each of which typically supports only a subset of these capabilities. The first feature is particularly distinctive: \textsc{PauliStrings.jl}~\cite{Loizeau_2025_Quantum,Loizeau_2025_Codebase,PauliStrings}, \textsc{PauliPropagation.jl}~\cite{rudolph2025pauli}, and the \textsc{spd} package~\cite{Begusic_2024_Fast,Begusic_2025_Real-time} all operate on finite lattices. While, for a given nesting order, the infinite-lattice result can be reproduced exactly using a finite lattice whose linear size is proportional to this order (as explained in  Section~\ref{subsec:translation}), this approach incurs a substantial memory overhead, which is critical since memory is the primary bottleneck in these computations. 

As for the symbolic treatment of Hamiltonian parameters, it is implemented in \textsc{PauliPropagation.jl}~\cite{rudolph2025pauli} and in the latest version of \textsc{PauliStrings.jl} \cite{PauliStrings,loizeau2026krylov}, but not in the \texttt{spd} package~\cite{Begusic_2024_Fast,Begusic_2025_Real-time}. Remarkably, retaining parameters in symbolic form incurs only moderate overhead in both memory and runtime compared to a purely numerical computation, owing to the inherently computer-algebraic nature of the algorithm. This feature is particularly advantageous when results are required for multiple points in parameter space: a single symbolic computation can be reused to obtain results for arbitrary parameter values \cite{rudolph2023classical,Uskov_Lychkovskiy_2024_Quantum,rudolph2025pauli,shirokov2025quench,ermakov2025symbolic}. Another, less obvious advantage of representing moments as polynomials in the parameters with integer coefficients, rather than as floating-point numbers, arises when the moments are subsequently processed within the recursion method. There, the symbolic representation helps to avoid numerical instabilities~\cite{Uskov_Lychkovskiy_2024_Quantum,Teretenkov_2024_Exact,shirokov2025quench} that otherwise plague the method \cite{Eckseler_2025_Escaping}.

Note that \textsc{QCommute} is intentionally limited to the computationally demanding task of generating nested commutators and extracting moments and quench coefficients from them. Further processing of these quantities is intentionally left to external codes. For example, within the recursion method, the moments are first converted into Lanczos coefficients and subsequently used to compute dynamical correlation functions and transport coefficients \cite{viswanath2008recursion,Parker_2019,Uskov_Lychkovskiy_2024_Quantum}. Since these post-processing steps are computationally inexpensive,  application-dependent, and continue to evolve together with the underlying methodology, we believe that keeping them outside the core package provides the most flexibility.

While \textsc{QCommute} has been developed with an emphasis on computational efficiency, flexibility is not currently its main strength. Enhancing flexibility is therefore a natural direction for future development. In particular, it would be desirable to extend the functionality of the code to support lattices beyond the hypercubic ones. Furthermore, it would be advantageous to implement the computation of quench coefficients for arbitrary initial states, rather than restricting to product states~\eqref{product state}. Beyond these largely technical improvements, a more substantial extension would be the direct implementation of the Pauli propagation algorithm~\cite{Begusic_2024_Fast,Begusic_2025_Real-time} within the package, fully leveraging its advanced parallelization and memory-management capabilities.


\section*{Acknowledgements}
We are grateful to Igor Ermakov for useful discussions and, in particular, for providing validation data for the three-dimensional Ising model.

\paragraph{Author contributions}
All three authors contributed to the development of the algorithm. OL initiated the project, developed a \textsc{Wolfram Mathematica}\textsuperscript{\textcopyright} code to validate the results for one-dimensional systems, and applied the results to physical problems. ISh implemented the algorithm in C++. VKh and ISh improved the performance of the code, with particular emphasis on parallelization and memory management. The manuscript was written by OL and ISh.

\paragraph{Funding information}
This work was supported by the Russian Science Foundation under grant \textnumero~24-22-00331 \url{https://rscf.ru/en/project/24-22-00331/}





\bibliography{SciPost_Example_BiBTeX_File.bib}


\end{document}